\documentclass[preprint,12pt]{elsarticle}

\usepackage{amssymb}
\usepackage{graphicx}
\usepackage[table,xcdraw]{xcolor}
\usepackage{acronym}
\usepackage{hyphenat}
\usepackage{hyperref}
\usepackage{courier}

\hypersetup{
    colorlinks=true,
    linkcolor=blue,
    }

\usepackage{lineno}

\journal{Engineering Applications of Artificial Intelligence}

\begin{document}

\begin{frontmatter}





\title{A Study on the Use of Edge TPUs for Eye Fundus Image Segmentation}



\author[label1]{Javier~Civit-Masot}
\author[label1]{Francisco~Luna-Perejón}
\author[label2]{José~María~Rodríguez~Corral}
\author[label1]{Manuel~Domínguez-Morales}
\author[label2]{Arturo~Morgado-Estévez}
\author[label1]{Anton Civit}

\address[label1]{Robotics and Computer Technology Lab. Computer Engineering Reseach Institute (I3US), Universidad de Sevilla, Seville, Spain.}
\address[label2]{School of Engineering, Universidad de Cádiz, 11519, Puerto Real, Spain.}

\begin{abstract}

Medical image segmentation can be implemented using Deep Learning methods with fast and efficient segmentation networks. Single-board computers (SBCs) are difficult to use to train deep networks due to their memory and processing limitations. Specific hardware such as Google's Edge TPU makes them  suitable for real time predictions using complex pre-trained networks. In this work, we study the performance of two SBCs, with and without hardware acceleration for fundus image segmentation, though the conclusions of this study can be applied to the segmentation by deep neural networks of other types of medical images. To test the benefits of hardware acceleration, we use networks and datasets from a previous published work and generalize them by testing with a dataset with ultrasound thyroid images. We measure prediction times in both SBCs and compare them with a cloud based TPU system. The results show the feasibility of Machine Learning accelerated SBCs for optic disc and cup segmentation obtaining times below 25 milliseconds per image using Edge TPUs.


\end{abstract}


\end{frontmatter}

\nolinenumbers

\section{Introduction}
\label{sec:introduction}

In recent years, the use of Deep Learning technologies for medical image analysis has quickly increased \cite{litjens2017survey,chen2020deep,teikari2019embedded,akkara2019role}. One of the main applications has been image segmentation, which is the process of detecting automatically or semi-automatically the limits within a two or three-dimensional image.

\begin{figure}[h]
  \centering
    \includegraphics[width=1\linewidth]{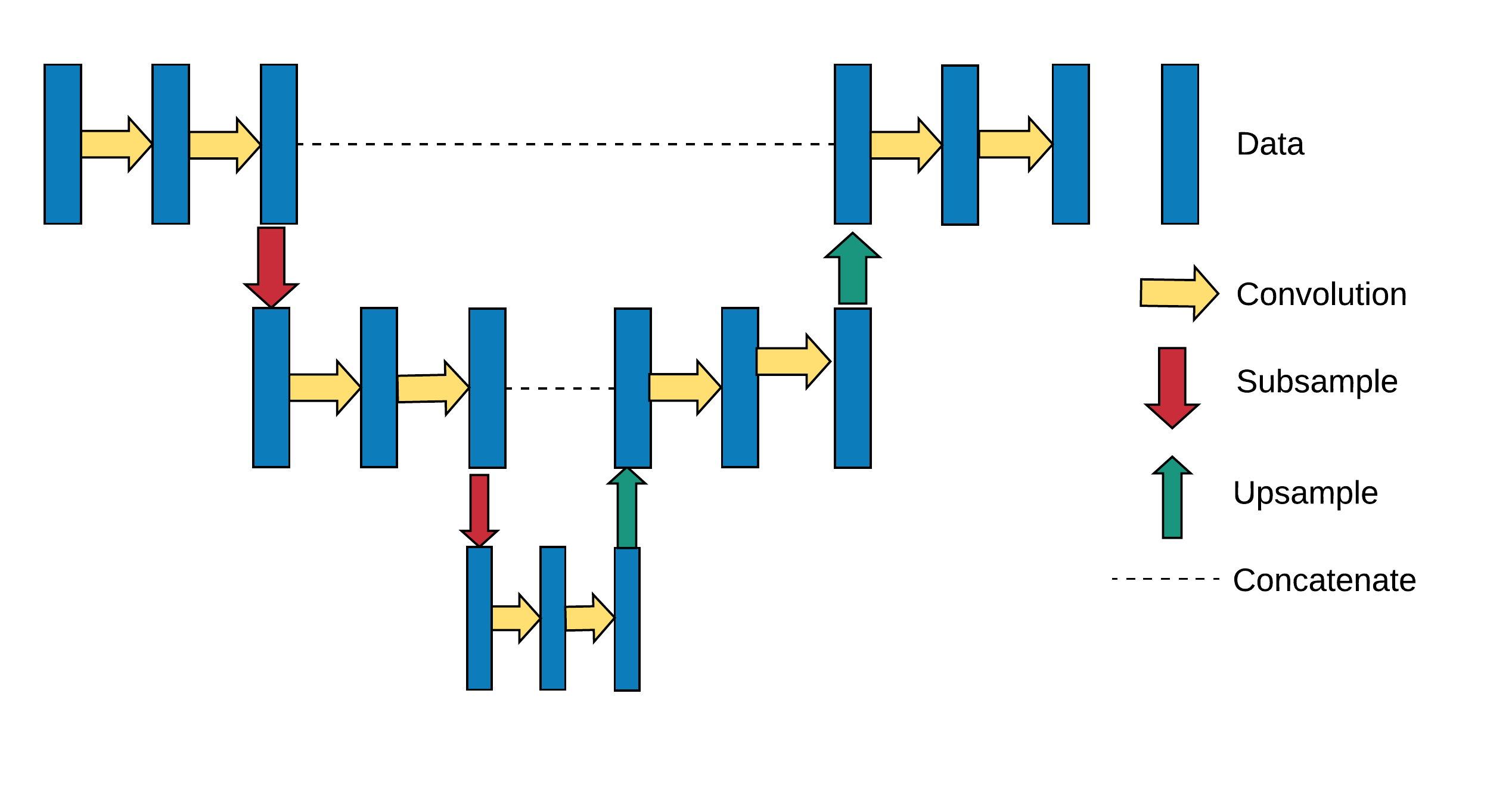}
  \caption{Basic three layer U-Net.}\label{fig:Unet}
\end{figure}

In  medical segmentation problems, many different segmentation networks have been used \cite{litjens2017survey}; however, a type of fully convolutional  neural network (CNN), known as U-Net \cite{ronneberger2015u}, has become very widely used and  shown to be very effective. U-Nets have been used with many types of medical images including X-Ray, MRI, CT, Ultrasound and eye Fundus images. The structure of a small three-layer generalized U-Net can be seen in Fig.  \ref{fig:Unet}. The network is made up of a set of descending layers, each with a larger number of filters but with the image resolution reduced to a quarter, an intermediate connecting layer (the bottom of the "U") and a set of ascending layers on which the original resolution is recovered.

Eye fundus images are widely used to help in Glaucoma detection. Glaucoma is a retinal  decease that can cause blindness in about 2\% and sight impairment in over 10\% of the cases \cite{quigley1985better}. Even though vision loss may occur even with optimum treatments, adequate therapy will stabilize the  majority of cases.

\begin{figure}
\centering
\includegraphics[width=5 cm]{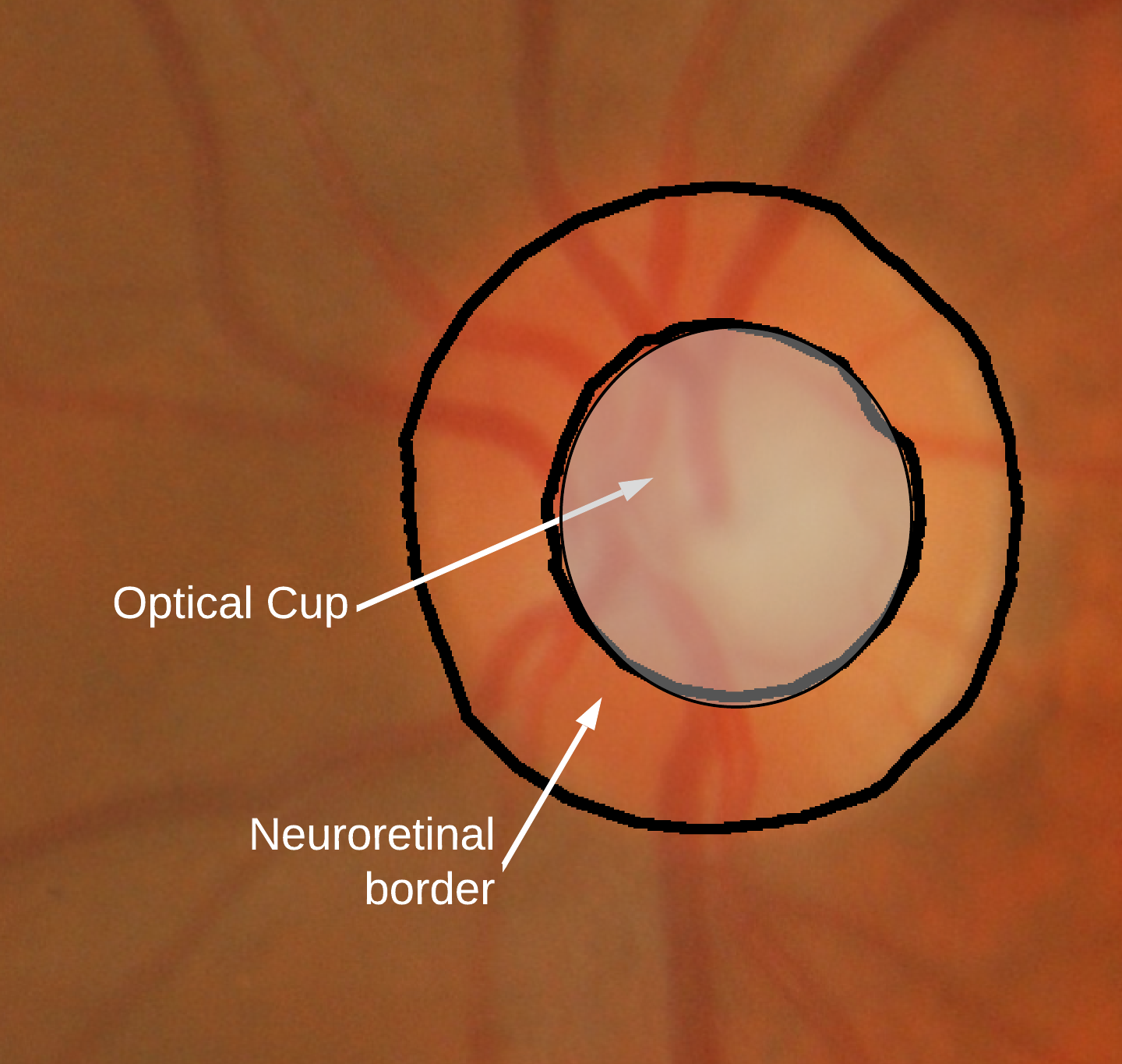}
\caption{Optic Disc and Cup}\label{optic_disc}
\end{figure}

The key to glaucoma detection is to understand how to examine the optic disc (OD) \cite{bourne2006optic}. The OD is an oval area where the retina connects to the optic nerve.  The optic cup (OC) is a white cup-like area in the center of the OD. The zone between the the cup and the disc is known as the neuroretinal rim. This region consists mostly of nerve fibers and is usually pink.  Most normal discs are mainly vertically oval with their cup  horizontally oval. A typical retina fundus image is shown in figure \ref{optic_disc}

Different indicators are used to help in diagnosing glaucoma from fundus images. The cup to disc ratio (CDR) \cite{maciver2017screening}, i.e. the relation between the diameters of the OD and the OC, is the most accepted glaucoma predictor. CDRs for glaucomatous and healthy eyes are about 0.65$$\pm$$0.13 and 0.39$$\pm$$0.15 respectively, thus establishing CDR as a valid diagnostic aid. An alternative detection method is based on the ISTN rule which uses the shape of the neuroretinal rim. On healthy eyes, the thickness of the  rim along the vertical and horizontal rim borders decreases in the order inferior (I)\textgreater superior (S)\textgreater nasal (N)\textgreater temporal (T) \cite{das2016diagnosis}. 

In our previous works \cite{civit2019tpu, civit2020dual}, generalized U-Nets were used for eye fundus image segmentation, specifically for optic disc (OD) and optic cup (OC) detection. The U-Net models were implemented on cloud-based GPU and TPU \cite{google2020cloud} architectures. 

An accurate OC and OD segmentation is important in order to calculate the CDR that, as already mentioned, is a well-established indicator for the diagnosis of glaucoma \cite{jonas2015optic,patel2018analysis,barros2020machine,cheng2013superpixel}.  The ISTN rule \cite{nath2012techniques} also relies on adequate OC and OD segmentation.

Currently, single-board computers (SBCs) have become popular as small low-cost computers for development, IoT and educational applications \cite{hassan2017internet,singh2017create}. The purpose of this work is to evaluate SBCs, with and without specific Machine Learning acceleration hardware, for implementing the generalized U-Net models developed in \cite{civit2020dual} for performing OD and OC segmentation.

In this work, we use very similar generalized U-Net architectures to segment  OD and  OC from fundus images and train them on Google TPUs. In particular, we use for cup segmentation a 6 level network with 64 channels in the first stage and a layer to layer filter increment ratio (IR) of 1.1. This is a lightweight model (2.5M trainable parameters), but as shown in \cite{civit2019tpu} it produces good results for cup segmentation. Although the model has an additional level compared to the original U-Net, and both have 64 channels in the first layer by decreasing the IR to 1.1 instead of the original 2.0, the number of parameters is reduced more than 50 times. For disk segmentation, as this is an easier problem, we use a similar network with only 40 channels in the first stage and less than 1M trainable parameters.

Even though we are using specific U-net examples, our aim is to show that embedded systems with specific Deep Learning acceleration hardware can perform many medical image segmentation problems in very reasonable times. This problem is very interesting in real medical practice as these hardware accelerated embedded processors can be easily included in lightweight portable medical instruments that can perform segmentation on their own without requiring an external PC. 

In this sense, segmentation by deep neural networks can also be performed in many other types of medical images  \cite{litjens2017survey} \cite{chen2020deep}, thus allowing this technology to be embedded in a wide range of portable medical diagnosis instruments.

Hence, the conclusions of  this work can be applied, for example, to the segmentation performed by Machine Learning accelerated embedded systems for organs and other substructures in cardiac or brain analysis, or for multi-organ segmentation (widely used for abdominal organ segmentation) \cite{Hesamian2019deep}.

In this work, more specifically, we implement the segmentation subsystem in \cite{civit2020dual} in two embedded devices. The first is a Raspberry Pi with no specific Machine Learning hardware and the second a Coral Dev Board with a Google's Edge TPU. Due to the limited resources of SBCs, even with specific accelerators, in terms of processing power and main memory size, it is technically unfeasible to train relatively large CNNs using them. However, they can be used for prediction purposes, as prediction is computationally a much less demanding process than training.

The Raspberry Pi has become very popular because of its affordable price, its ease of use due to the availability of the Raspbian operating system (a derivative of Debian Linux), and its large development community. Moreover, the new Raspberry Pi 4 includes a more powerful processor and up to 4 GB of RAM. Thus we use this very well known device as one of our reference systems to be able to compare its results with  devices that implement hardware acceleration  for Machine Learning.

The Coral Dev Board is a SBC specifically designed to perform Machine Learning inferencing in a small-form factor. It includes a simplified tensor processing unit (TPU), the Edge TPU, which is an ASIC developed by Google for providing high performance Machine Learning inferencing with a low power usage. A very similar SBC, also with Edge TPU, is the Tinker Edge T board from Asus. 

The purpose of this work consists not only of confirming the feasibility of implementing the U-Nets used in \cite{civit2020dual} to perform OD and OC segmentation in the mentioned  SBCs, but also finding if they can make predictions in a reasonable time. We also want to compare these prediction times with those obtained using cloud-based GPU and TPU devices.

As already discussed, it would be interesting and convenient for an ophthalmologist to be able to segment and obtain assistance for his or her diagnostic, directly from the image acquisition medical instrument. This would avoid the need of using a local GPU PC or having to upload the images to the web.

Thus, the importance and  usefulness of SBCs with Deep Learning capabilities lies in the fact that these devices can operate autonomously. They do not need to be connected to servers equipped with GPUs or TPUs for performing predictions, since they have Machine Learning hardware built-in.

Moreover, Cloud GPUs and TPUs are usually free for non-commercial uses only. Therefore, the use of SBCs provided with Machine Learning hardware, such as the Coral Dev Board (that includes a Google Edge TPU in its design) or the NVIDIA Jetson boards\footnote{\href{https://developer.nvidia.com/embedded/jetson-developer-kits}{https://developer.nvidia.com/embedded/jetson-developer-kits}} (that are equipped with GPUs), are interesting options to consider.

Also, the use of a Machine Learning accelerated SBC for performing medical image segmentation ensures the privacy protection of patient’s health data, since the information is used locally by the SBC and, thus, it not sent to any cloud server for processing.

The rest of the paper is structured as follows: The background and related works are presented in Section 2. The methodological aspects are explained in Section 3, in order to describe the design of the experimental tests and thus provide a better understanding of the process followed to obtain the results. These results are described in Section 4 and discussed in Section 5. Finally, Section 6 draws the conclusions of this work and proposes future research lines.

\section{Background and Related Works}
\label{sec:back_and_related}

A study of generalized U-Net architectures was performed in \cite{civit2019tpu} as a technique for implementing eye fundus image segmentation in the cloud. 

In \cite{civit2019tpu} U-Net implementations deeper than the standard 5-layer network and with different layer increment ratios were tested. The use of normalization and drop-out as well as the influence of the initial layer width and the layer to layer width ratio (IR) were studied, since these attributes affect significantly both prediction quality and learning speed, and vary widely among different implementations.

Moreover, in a cloud based scenario, the neural networks must be trained as independently as possible from the acquisition source, since in a cloud-based service images will come from very different sources. Several segmentation researchers have used various datasets, but they always train and test with each of them independently. In our implementation  data from several datasets were preprocessed and mixed in order to create independent datasets for training and validation.

Publicly available datasets were used. DRISHTI-GS \cite{sivaswamy2014drishti}, from Aravind Eye Hospital, Madurai (India), is a set of fundus images labeled by expert ophthalmologists for disc and cup. RIM-ONE-v3 \cite{fumero2011rim}, from the MIAG group at the University of La Laguna (Spain), is a set of fundus images also labeled for disc and cup. Finally, DRIONS-DB \cite{carmona2008identification}, from Miguel Servet Hospital, Saragossa (Spain), is a set of fundus images where only the optic cup has been labeled.

As a result of that initial study, a set of functions that enable the implementation of generalized U-Nets adapted to TPU execution was developed. These U-Nets are also suitable for developing cloud-based service implementations.

Regarding the use of embedded platforms and single-board computers for  medical image segmentation, a low-cost Deep Learning ready GPU embedded platform has been used in \cite{ Niepceron2020Moving} for performing segmentation of brain tumors. An Nvidia Jetson AGX Xavier was selected in this study due to its low weight and low power consumption characteristics. Also, this developer kit  embeds a modular scalable architecture called Deep Learning Accelerator which includes a support for many widely used  CNNs.

An existing Deep learning architecture was selected and modified to be usable for both training and inference on the Jetson AGX Xavier platform. More specifically, a MobileNetV2 architecture has been compressed to reduce the number of trainable parameters and increase the training speed. Also, neural network 8-bit fixed-point quantization has been used for performing inferences in addition to the compression of the convolution layers.

Using the Jetson AGX Xavier maximum capacity, the compressed and quantized model was successfully trained, and it was able to segment high and low grade gliomas. The authors compared the model performance with other state of the art  approaches, and proved that their method reached comparable results in relation to the reduction of parameters.

More specifically,  regarding the use of embedded systems, mobile devices and single-board computers for performing eye fundus segmentation, a HW/SW embedded system that implements a Vertical Cup-to-Disc Ratio (VCDR) evaluation method for the diagnosis of glaucoma was presented in \cite{dantas2016hw}. This method, which is based mainly on morphological operations, has a reasonably low computational cost, but maintains an accuracy comparable to other related works using the RIM-ONE dataset.

It can be implemented on low power embedded processors with FPGA-based hardware acceleration for the morphological operations, to reduce execution time while maintaining accuracy. The proposed FPGA assisted architecture  reduces execution time by at least 30\% compared with software-only implementations running on platforms based on low power processors such as Raspberry Pi Model B, BeagleBoard-xM and a system using an Intel Atom processor with 2 GB DRAM.

In \cite{martins2020offline}, an interpretable computer-aided diagnosis (CAD) pipeline, that runs offline in mobile devices, is used for diagnosing glaucoma using fundus images. Several  public datasets were merged and used to train convolutional networks for performing classification and segmentation tasks.

These networks were  used to build a pipeline that outputs a glaucoma confidence level, and also provides several morphological features and segmentations of relevant structures, resulting in an interpretable diagnosis in a similar fashion to \cite{civit2020dual}. This pipeline was integrated into a mobile app that  run on  a Samsung Galaxy S8 smartphone. Execution times - for CPU and GPU - and memory requirements were assessed.

In a similar way, in \cite{perez2020lightweight} a Deep Learning method for assessing the eye fundus image quality small enough to be deployed in a smartphone was presented. This method was validated with two different datasets, achieving good  accuracy results.

The authors also measured the classification average elapsed time for the binary and three-class models on a smartphone running Android 9.0. The proposed method has a small number of parameters in comparison with other state-of-the-art models, and, thus, it is an attractive alternative for a mobile-based eye fundus quality classification system.

\section{Materials and Methods}
\label{sec:mat_met}

We will start this section with a brief description of the hardware resources used. Then, the specification of the parameters of the generalized U-Nets selected for OC and OD segmentation as well as the  datasets used, will complete the design of the experimental tests.

\subsection{Hardware}

\begin{figure}[h]
  \centering
    \includegraphics[width=.50\linewidth]{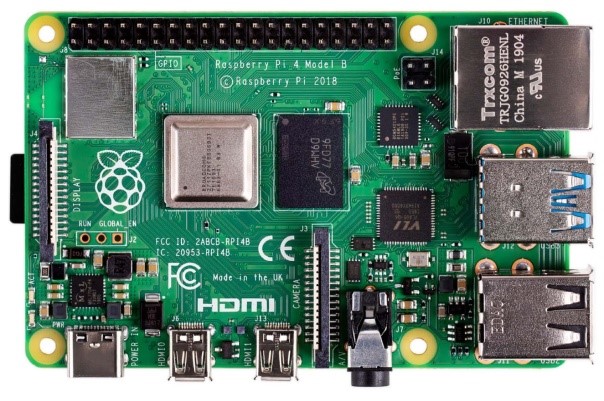}
  \caption{Raspberry Pi 4 Model B.}\label{fig:PI}
\end{figure}

Raspberry Pi\footnote{\href{https://www.raspberrypi.org/}{https://www.raspberrypi.org/}} can be considered as a general-purpose computing device (Fig. \ref{fig:PI}), usually with a Linux operating system, that can run multiple programs in a multitasking environment. The Broadcom system-on-chip BCM2711 used by the latest version of the board (Raspberry Pi 4 Model B) \cite{halfacree2018official} includes in its design a 64-bit quad-core Cortex-A72 ARM processor @ 1.5 GHz along with the new VideoCore VI 3D unit, and also a natively attached Gigabit Ethernet controller as well as a PCIe link that connects the USB ports.

Raspberry Pi 4 is also capable of addressing 1 GB, 2 GB or 4 GB LPDDR4 RAM depending on the variant of the model. The on-board wireless LAN (dual-band 802.11 b/g/n/ac) and Bluetooth 5.0 low-energy (BLE) connection capabilities make this device useful for the development of IoT applications. This single-board computer can run a variety of operating systems, such as Raspbian, which is the Foundation's officially supported operating system, Ubuntu Mate and Windows 10 IoT Core. We include this device in order to verify the performance of a widely used non hardware accelerated SBC in medical segmentation applications.

\begin{figure}[h]
  \centering
    \includegraphics[width=.50\linewidth]{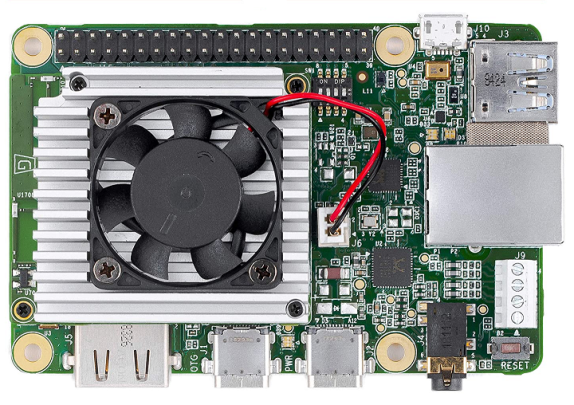}
  \caption{Coral Dev Board.}\label{fig:Coral}
\end{figure}

The Coral Dev Board \cite{coral2020datasheet} is a single-board computer specifically developed to perform Machine Learning inferencing (Fig. \ref{fig:Coral}). The Coral System-on-module (SoM), which is part of this prototype board, can be bought separately and used in a custom PCB hardware for production purposes.

The SoM is an integrated system that includes the NXP’s IMX 8M system-on-chip (Quad-core Arm Cortex-A53 @ 1.5 GHz and Arm Cortex-M4F processors plus a Vivante GC7000Lite GPU), 8 GB eMMC memory, 1 GB LPDDR4 RAM, Wi-Fi 802.11 b/g/n/ac and Bluetooth 4.2 connection capabilities, and a Google Edge TPU coprocessor (a small ASIC which provides high performance Machine Learning inferencing for TensorFlow Lite models).

The baseboard for the SoM includes the usual connectors needed to perform a prototype project, such as Gigabit Ethernet port, CSI-2 camera interface, DSI display interface, 40 I/O pin header and USB 2.0/3.0 ports. Coral Dev Board uses Mendel operating system, which is a lightweight derivative of Debian Linux that runs on several Coral development boards.

\subsection{Datasets}

Regarding the datasets used in this work, both DRISHTI-GS and RIM-ONE-v3 have been used. Both are publicly available datasets and provide human expert OC and OD segmentation data and additional labels indicating if the images correspond to a patient with glaucoma or not. The labeling process includes the supervised evaluation of each of the dataset samples by a professional in the field.

There are other datasets, such as DRIONS-DB, that was used in previous studies, but it is not used in this work because only the optic cup has been labeled. However, this is not necessarily a drawback for achieving the objective of our study, specified in the introduction section, since relevant results can be obtained using the other two datasets.

DRISTI-GS dataset from Aravind Eye Hospital, Madurai (India), is made up of 101 color fundus images labeled for both disc and cup; and RIM-ONE dataset from the University of La Laguna is composed of 159 images also labeled for disc and cup.

In this work, 75\% of the images from each dataset is used for training and the remaining 25\% of the images for validating the results. This can be observed in Table \ref{tab:dataset}.

\begin{table}[h!]
    \centering
    \caption{Dataset summary}
    \label{tab:dataset}
    \begin{tabular}{c|cccc}
        \hline
        Dataset & Images &
        \begin{tabular}{c}
            Images\\
            after D.A.
        \end{tabular}
         
        & Train (75\%) & Test (25\%)\\
        \hline
        DRISHTI-GS &101 &2380 &1785 &595\\
        RIM-ONE-v3 &159 &6980 &5235 &1745\\
        TOTAL &260 &9360 &7020 &2340\\
        \hline
    \end{tabular}%
\end{table}

The first column shows the number of images that are provided in those public datasets, the second column indicates the final amount of images used after data augmentation processes and, finally, the other two columns  present the number of images used for training and testing purposes respectively.

Finally, in order to have more available data and thus to obtain more experimental results, a new dataset of thyroid gland ultrasound images \cite{wunderling2017comparison} has been included. Thus, our study is extended to the segmentation of medical images different from those of eye fundus, that are obtained by other acquisition methods. This new dataset consists of 3665 images with their respective labels.

\subsection{System architecture and testing method}

After describing the hardware resources and datasets used, we will address the other aspects concerning the system architecture and the design of the experimental tests. In relation to the U-Nets, we have selected from \cite{civit2019tpu} a network with 6 levels, 40 filters in the initial layer and a layer-to-layer increment ratio of 1.1 (identified as 6/40/Y/1.1) with 0.9 MTP (millions of trainable parameters) for optic disc segmentation. We have selected a network with 6 levels, 64 filters in the initial layer and a layer-to-layer increment ratio of 1.1 (identified as 6/64/Y/1.1) with 2.4 MTP for optic cup segmentation. The small IR value reduces the number of trainable parameters greatly and is the key to efficient embedded implementation.

The first network is one of the U-Nets that provided better results for the optic disc segmentation. As for the optic cup segmentation, we have selected a U-Net used in \cite{civit2020dual} that, without providing the best results (though they can be considered very good also), allows the generation of a suitable Tensor-Flow Lite model adequate to be processed by the Edge TPU Compiler. Logically, this tool does not admit models whose information does not fit in the memory size of the Edge TPU coprocessor.

\begin{figure}[h]
  \centering
    \includegraphics[width=1.0\linewidth]{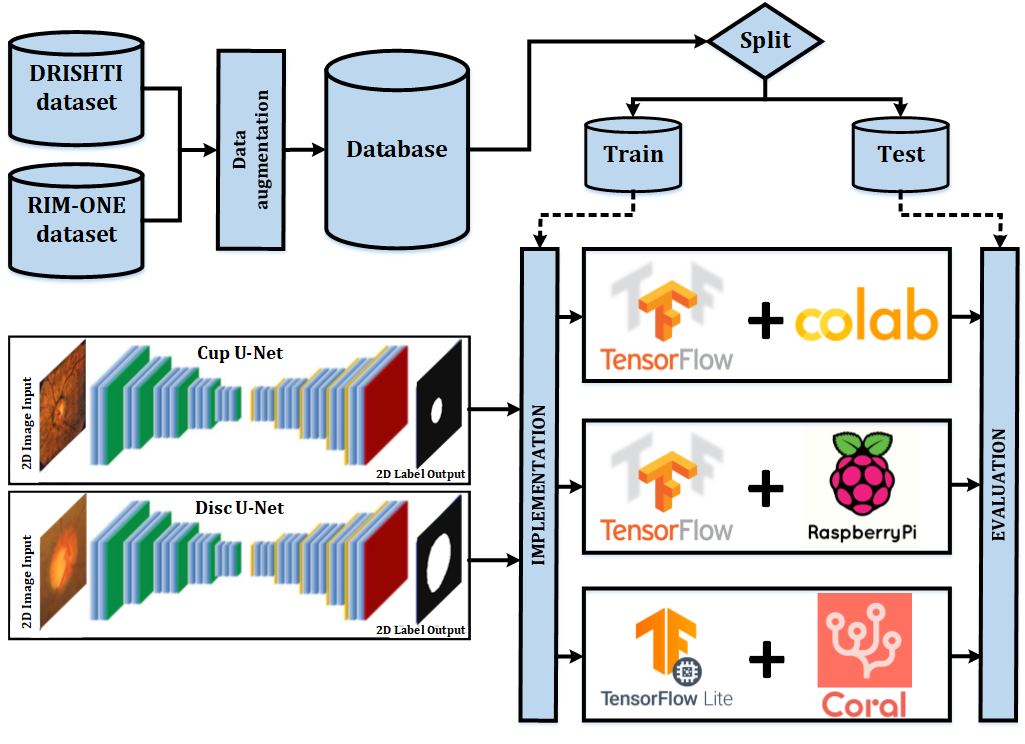}
  \caption{Full system implementation.}\label{fig:GU}
\end{figure}

A global graphical abstract of the  implemented and tested system is shown in Fig. \ref{fig:GU}.

\begin{figure}[h]
  \centering
    \includegraphics[width=.50\linewidth]{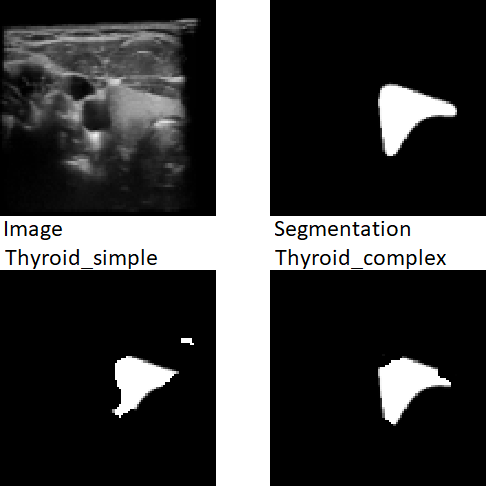}
  \caption{Images from thyroid, segmentation, and predictions obtained with Thyroid\_simple and Thyroid\_complex models.}\label{fig:functThyr}
\end{figure}

Additionally, these two U-Nets have been retrained using the thyroid ultrasoud dataset in order to obtain two new models for our study. The first model (Thyroid\_simple) has been obtained by retraining the U-Net used for OD segmentation, and the second one (Thyroid\_complex) has been obtained by retraining the more complex U-Net used for performing OC segmentation (Fig. \ref{fig:functThyr}).

For all the experimental tests, prediction times have been obtained calling the \texttt{timeit.default\_timer()} Python method just before and after making a prediction with the specific model.

Next, we show a general scheme of the developed test programs:
\begin{itemize}
    \item Image dataset load.
    \item TensorFlow model definition and compilation, and load of the model weights for performing the experimental tests using the Raspberry Pi board and the iPython notebooks.
    \item Alternatively, TensorFlow Lite model conversion and load when using the Coral Dev Board for performing the experimental tests.
    \item Prediction execution with time measurement.
\end{itemize}

\section{Results}
\label{sec:results}

 The segmentation performance of the proposed system has already been studied in \cite{civit2019tpu} and \cite{civit2020dual}.  For completeness we include in Table \ref{tab:segDice} the Dice coefficients \cite{sorensen1948method} obtained by our approach compared with other Deep Learning based alternatives that use the same public datasets. As we can see, results are fully comparable with other works even though our networks are trimmed for embedded implementation and we train with a combined dataset while the remaining authors train specifically for each dataset.
 
 In \cite{civit2020dual} we used the same network to segment the disk and cup while in this work we decided to use an  smaller network for cup segmentation. This reduces the number of trainable parameters by a factor of almost 3, but clearly has some impact in the segmentation performance. If we used the same network for both OC and OD segmentation the Dice coefficient would be the same as those in \cite{civit2020dual}.
 
 Regarding the segmentation performance, another aspect that needs to be mentioned is that on edge TPUs the models have to be quantized to 8 bit fixed precision numbers. We can see that this has a small impact on the segmentation performance. The obtained Dice coefficients when running the quantized models on Edge TPUs are shown on the last row of Table \ref{tab:segDice}.
 
 In the rest of the paper we will include results only related to prediction times, as this is the main objective of this work and the segmentation performance is almost identical in the different proposed implementations.

\begin{table*}[h]
    \centering
    \caption{Dice Coefficients for Cup and Disc segmentation}
    \label{tab:segDice}
\resizebox{0.98\textwidth}{!}{
\begin{tabular}{l|cccc}

\hline
Author                                                  &{Cup RIM-ONE} & Disc RIM-ONE & Cup DRISHTI & Disc DRISHTI \\ \hline

Zilly \cite{Zilly2017glaucoma}                  & {-}           & -            & 0.87          & 0.97           \\
Sevastopolsky  \cite{sevastopolsky2017optic}                           & 0.82                               & 0.94           & -           & -            \\

Shankaranarayana \cite{Shankaranarayana2017disc}  & 0.94                               & 0.98           & -           & -            \\
Al-Bander \cite{albander2018dense}                                           & 0.69                               & 0.90           & 0.83          & 0.95           \\
Civit-Masot \cite{civit2020dual}                                &0.84                                  &0.92             &0.89             &0.93   \\
CPU/GPU/TPU &0.84                                  &0.86            &0.89             &0.91   \\
Edge TPU\textsuperscript{*}  &0.84                                  &0.85             &0.88             &0.90  

\\ \hline

\end{tabular}
}
\\
\resizebox{0.45\textwidth}{!}{
\begin{tabular}{l}
\textsuperscript{*} \textit{Results obtained with quantized models} 
\end{tabular}
}

\end{table*}

First, we have obtained a set of prediction times with the two selected U-Nets using Google Colaboratory notebooks\footnote{\href{https://colab.research.google.com}{https://colab.research.google.com}}. These times have been obtained for GPU, CPU and TPU to be used as reference values (Tables \ref{tab:tab1} and \ref{tab:tab2}), so that we can compare them with those prediction times obtained with Raspberry Pi and Coral Dev Board SBCs.

As in \cite{civit2019tpu}, we have used the Google Colaboratory iPython notebook development environment. This environment supports TensorFlow and Keras \cite{chollet2018deep}, and allows the implementation and training of networks using GPUs and TPUs \cite{google2020cloud} in Google Cloud. In order to obtain prediction times in Google Colaboratory environment, we have used 2.4.1 and 2.4.0 versions of TensorFlow and Keras respectively.

Predictions using Colab notebooks have been made on an Intel(R) Xeon(R) CPU @ 2.20 GHz using a single core with two threads \cite{google2020colab}. An Nvidia Tesla T4 has been used for making predictions using GPUs. Also, for performing predictions using Google Cloud TPUs, v2 TPU Pods have been used. A TPU v2 has 8 GiB of high-bandwidth memory and one matrix unit (MXU) for each TPU core. A v2 TPU Pod is a cluster consisting of up to 512 TPU cores and 4 TB of total memory \cite{google2020cloud}.

The first part of Tables \ref{tab:tab1} and \ref{tab:tab2} shows the results obtained using Google Colaboratory environment for optic disc and cup, and thyroid segmentation. These times have been calculated from the next ten predictions on a dataset after performing the initial prediction. Since TensorFlow performs a prediction over an entire dataset, the prediction time for a single element can be calculated as the total prediction time for a dataset divided by its number of elements. Thus, once the ten image prediction times have been obtained, the mean prediction  time per image along with its standard deviation is shown.

The first prediction on a dataset using CPU and GPU takes more time than the next ones due to the necessity of performing a set of memory allocations and initializations. In the case of prediction times using Cloud TPU, the first prediction also involves sending the model information through the network and copying it into the TPU memory.

For implementing the U-Nets in Raspberry Pi 4 Model B and performing the experimental tests, 2.1.0 and 2.2.4-tf versions of TensorFlow and Keras \cite{qengineering2020install} have been used respectively. As with previous results and due to the same reason stated before, the corresponding times for Raspberry Pi using CPU have also been calculated using the next ten predictions on a dataset after the first inference. Again, once the ten image prediction times have been obtained dividing each total prediction time for a dataset by its number of elements, the mean prediction  time per image along with their standard deviation are also shown.

For performing the experimental tests with Coral Dev Board, we have used TensorFlow Lite 2.5.0, which is the framework used to perform Deep Learning inferences on the Edge TPU \cite{google2020start}. First, predictions on DRISHTI-GS and RIM-ONE datasets with the TensorFlow Lite models for OD and OC segmentation, and predictions on THYROID dataset with the TensorFlow Lite models for segmenting the thyroid (Thyroid\_simple and Thyroid\_complex), have been obtained using the Coral Dev Board CPU.

In order to adapt the original TensorFlow models for the two U-Nets to the format used by TensorFlow Lite, the post-training quantization technique \cite{tensor2020post,google2019retrain} has been used. The next step is to obtain suitable versions of the TensorFlow Lite models to be executed on the Edge TPU coprocessor \cite{google2020tensor}. The Edge TPU Compiler \cite{google2020edge} process the information of a TensorFlow Lite model and generates an Edge TPU compatible model.

\begin{table}[h]
    \centering
    \resizebox{0.99\textwidth}{!}{
        \begin{tabular}{c|ccc}
            \hline
             &  Google Colaboratory & Raspberry Pi 4B & Coral Dev Board\\
             \hline
             \begin{tabular}{c}
                  Dataset (shape)   \\
                  DRISHTI (595, 128, 128, 3) \\
                  RIM-ONE (1745, 128, 128, 3) \\
                 THYROID (3665, 128, 128, 3)
             \end{tabular}
              & 
             \begin{tabular}{ccc}
                 \begin{tabular}{c}
                      CPU   \\
                      73.11$\pm${0.44} \\
                      71.84$\pm${0.12} \\
                     76.22$\pm${0.09}
                 \end{tabular}
                &
                 \begin{tabular}{c}
                      GPU   \\
                      2.16$\pm${0.05} \\
                      1.59$\pm${0.04} \\
                      1.42$\pm${0.01}
                 \end{tabular}
                &
                 \begin{tabular}{c}
                      TPU   \\
                     17.24$\pm${1.91} \\
                      7.71$\pm${1.19} \\
                      4.88$\pm${0.37}
                 \end{tabular}
             \end{tabular}
             &
             \begin{tabular}{c}
                  CPU   \\
                  259.52$\pm${0.52} \\
                  256.15$\pm${0.59} \\
                 255.35$\pm${0.26}
             \end{tabular}
             & 
             \begin{tabular}{cc}
                 \begin{tabular}{c}
                      CPU   \\
                      576.45$\pm${0.19} \\
                      576.37$\pm${0.89} \\
                      575.71$\pm${0.73}
                 \end{tabular}
                &
                \begin{tabular}{c}
                    TPU   \\
                    8.55$\pm${0.90} \\
                    8.73$\pm${1.16} \\
                   8.31$\pm${1.06}
                \end{tabular}
             \end{tabular}
             \\
             \hline
        \end{tabular}
    }
    \caption{Image prediction times for Optic Disc and Thyroid\_simple (in milliseconds)}
    \label{tab:tab1}
\end{table}

\begin{table}[h]
    \centering
    \resizebox{0.99\textwidth}{!}{
        \begin{tabular}{c|ccc}
            \hline
             &  Google Colaboratory & Raspberry Pi 4B & Coral Dev Board\\
             \hline
             \begin{tabular}{c}
                  Dataset (shape)   \\
                  DRISHTI (595, 128, 128, 3) \\
                  RIM-ONE (1745, 128, 128, 3) \\
                  THYROID (3665, 128, 128, 3)
             \end{tabular}
              & 
             \begin{tabular}{ccc}
                 \begin{tabular}{c}
                      CPU   \\
                     163.72$\pm${1.41} \\
                     160.61$\pm${0.72} \\
                     155.07$\pm${2.96}
                 \end{tabular}
                &
                 \begin{tabular}{c}
                      GPU   \\
                      4.20$\pm${0.18} \\
                      2.92$\pm${0.21} \\
                      2.29$\pm${0.01}
                 \end{tabular}
                &
                 \begin{tabular}{c}
                      TPU   \\
                      38.49$\pm${1.43} \\
                      15.81$\pm${2.05} \\
                      4.81$\pm${0.15}
                 \end{tabular}
             \end{tabular}
             &
             \begin{tabular}{c}
                  CPU   \\
                  591.11$\pm${1.21} \\
                  581.56$\pm${1.01} \\
                  576.83$\pm${0.97}
             \end{tabular}
             & 
             \begin{tabular}{cc}
                 \begin{tabular}{c}
                      CPU   \\
                      1148.76$\pm${0.22} \\
                      1149.64$\pm${1.65} \\
                     1145.45$\pm${0.54}
                 \end{tabular}
                &
                \begin{tabular}{c}
                    TPU   \\
                    21.64$\pm${0.95} \\
                    21.47$\pm${1.14} \\
                    21.87$\pm${0.81}
                \end{tabular}
             \end{tabular}
             \\
             \hline
        \end{tabular}
    }
    \caption{Image prediction times for Optic Cup and Thyroid\_complex (in milliseconds)}
    \label{tab:tab2}
\end{table}

\begin{figure}[h]
  \centering
    \includegraphics[width=.50\linewidth]{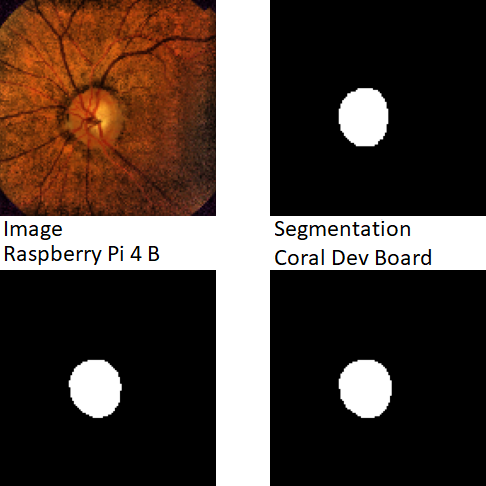}
  \caption{Images from optic disc, segmentation, and predictions obtained with Raspberry Pi and Coral Dev Board.}\label{fig:functDisc}
\end{figure}

\begin{figure}[h]
  \centering
    \includegraphics[width=.50\linewidth]{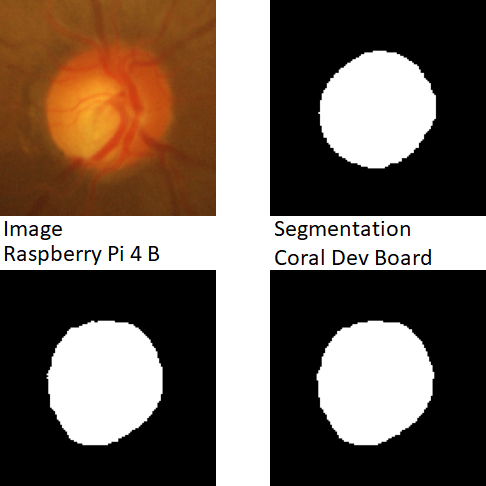}
  \caption{Images from optic cup, segmentation, and predictions obtained with Raspberry Pi and Coral Dev Board.}\label{fig:functCup}
\end{figure}

At the end of the process, this tool also generates a “.log” file indicating which operations of the original TensorFlow Lite model have been mapped to the Edge TPU coprocessor, and which operations will continue running on the Coral Dev Board CPU.

Moreover, the inference program that includes the model compiled for the Edge TPU must use a TensorFlow Lite delegate \cite{google2020run}, so that whenever the interpreter finds a graph node compiled for the Edge TPU, it sends that operation to the coprocessor executing the rest of the program on the ARM CPU. Thus, the file name of the Edge TPU runtime library must be passed to the \texttt{load\_delegate()} method when initializing the interpreter.

The last part of Tables \ref{tab:tab1} and \ref{tab:tab2} shows the results for Coral Dev Board. Inference programs using TensorFlow Lite models do not perform predictions on an entire dataset as TensorFlow programs do (by calling the \texttt{predict()} method). Instead, TensorFlow Lite programs perform their predictions on the individual elements of a dataset. Thus, a loop must be used for iterating over each element.

Moreover, the first prediction time is not considered, since the first inference on the Edge TPU coprocessor is slow as it includes the load of the TensorFlow Lite model into the memory of the device \cite{google2020start}. Therefore, after performing the first inference, the prediction loop starts iterating over the first image of the dataset so that the first inference is performed again.

Finally, Figs. \ref{fig:functDisc} and \ref{fig:functCup} show images from the optic disc and cup respectively along with the corresponding segmentation made by ophthalmologists, as well as the predictions performed with Raspberry Pi 4 Model B and Coral Dev Board SBCs.

\section{Discussion}
\label{sec:discussion}

Once we know that the two SBCs selected for our study - Raspberry Pi and Coral Dev Board - are valid for performing segmentation of eye fundus and thyroid ultrasound images, and that predictions performed by these devices are basically equal to predictions made by Cloud CPUs, GPUs and TPUs when using Google Colaboratory notebooks, it is necessary to evaluate the performance of such devices - in particular the Coral Dev Board as it is equipped with specific Machine Learning hardware - and thus verify if they are capable of making predictions in reasonable times.

Prediction times for CPU, GPU and TPU using Google Colab notebooks are clearly smaller than those obtained for Raspberry Pi 4 Model B, as expected. This result can be easily explained, since the performance of the CPU used in Colab notebooks (Intel(R) Xeon(R) CPU @ 2.20 GHz) is greater than that of Raspberry Pi 4 Model B CPU (Cortex-A72 ARM processor @ 1.5 GHz). Moreover, prediction times obtained when using Colab notebooks are obviously smaller for GPU and TPU than the ones obtained for CPU.

However, prediction times for TPU are greater, to some extent, than prediction times for GPU. This result can be explained by the delays due to data transmission over a network \cite{diaz2016extending}, since the CPU and the GPU are in the same node (and the communication between them is local) but the TPU pod \cite{google2020cloud} is in another node. This makes cloud based TPUs much more useful for network training than for predictions on small data samples.

Regarding the results  for Coral Dev Board, although our interest  focuses on prediction times for the Edge TPU coprocessor, we also show prediction times for CPU in Tables \ref{tab:tab1} and \ref{tab:tab2}. Thus, we can use them as reference values for highlighting the performance improvement when making predictions using Edge TPUs.

The corresponding results for optic disc and cup and thyroid (Tables \ref{tab:tab1} and \ref{tab:tab2}) show that prediction times for the TensorFlow Lite models are appreciably greater than those for the TensorFlow models when using the Raspberry Pi 4 Model B board. This can be explained by the higher performance of Raspberry Pi 4 Model B CPU (Cortex-A72 @ 1.5 GHz) in relation to that of Coral Dev Board CPU (Cortex-A53 @ 1.5 GHz). Cortex-A72 microprocessor supports out-of-order execution, has a 15-stage pipeline (against the 8-stage pipeline of Cortex-A53), a more sophisticated branch predictor and greater L1 and L2 caches.

Regarding prediction times for Edge TPU, which are our main objective, we can see a significant performance improvement, as expected, in comparison with prediction times for CPU in Google Colab notebooks, Raspberry Pi and Coral Dev Board itself.

Finally, when comparing TPU prediction times, it can be observed that with a sufficiently large dataset (as RIM-ONE-v3 and THYROID in this case), prediction times in Tables \ref{tab:tab1} and \ref{tab:tab2} are smaller for Colab notebooks compared with prediction times for Coral Dev Board. The opposite is the case when the number of elements in the dataset is relatively small (as with DRISHTI-GS).

Since the CPU for a Colab notebook and the cluster of TPUs (TPU Pod) \cite{google2020cloud} are in different nodes of a network, there is a data transmission time \cite{diaz2016extending} which can be considered to be bounded except for a technical incidence that may araise in the system.

Thus, when the number of dataset images on which the predictions are performed is relatively large, the data transmission time ceases to be significant in relation to the total time for the set of predictions performed by Cloud TPUs, which have a much greater performance than the Edge TPU coprocessor. This last device is primarily intended for model inferencing, but not for training large and complex Machine Learning models\footnote{\href{https://coral.ai/docs/edgetpu/faq/}{https://coral.ai/docs/edgetpu/faq/}} which is one of the main objectives of cloud based TPUs.

Therefore, for a sufficiently large number of predictions (n) on a dataset, the total time using Cloud TPUs along with the data transmission time over the network ends up being smaller than the total time used for predictions on the dataset using the Edge TPU.

\begin{equation}
  SUCTET = \lim_{n \to \infty} \frac{n \times ETPT}{NDTT + n \times CTPT}\\
  \label{eqn:su1}
\end{equation}

The speed up (SUCTET) in the performance of Cloud TPUs for Colab notebooks versus Edge TPU for Coral Dev Board can be expressed with equation \ref{eqn:su1}. When the number of images (n) in the dataset on which predictions are made is relatively large, the network data transmission time (NDTT) ceases to be significant in relation to the total time for the set of inferences performed by Cloud TPUs. The term CTPT indicates the prediction time on a dataset element for the Cloud TPU pod, whereas ETPT indicates the prediction time for the Edge TPU on an element of the same dataset.

\begin{equation}
  n \times ETPT < NDTT + n \times CTPT\\
  \label{eqn:su2}
\end{equation}

\begin{equation}
  n < \frac{NDTT}{ETPT - CTPT}\\
  \label{eqn:su3}
\end{equation}

In order to know the maximum value for the number of images (n) of a dataset for which the performance of the Coral Dev Board Edge TPU is better than the performance of the Cloud TPU, the total prediction time on this dataset must be smaller when using the Edge TPU (equation \ref{eqn:su2}). Thus, for values of n that are lower than a maximum limit (equation \ref{eqn:su3}), the total prediction time using the Edge TPU will be smaller than the total prediction time using the Cloud TPU.

\section{Conclusions and Future Works}
\label{sec:conclusions}

The feasibility of using single-board computers for segmenting eye fundus images with a Deep Learning model in a reasonable time has been demonstrated  experimentally: Less than 1.2 seconds per image for the worst case (using Coral Dev Board CPU) and less than 9 milliseconds per image for the best case (using Coral Dev Board Edge TPU). It is clear that including specific Machine Learning hardware accelerators provides an speedup of over 130 times and thus allows many sophisticated segmentation problems to be performed in real time on embedded devices, such as many medical image acquisition instruments.

As future work, we plan to extend our study by including not only segmentation but also direct Glaucoma classification subsystems to be able to build explainable glaucoma diagnosis aids directly on the acquisition instrument. We also plan to use Coral accelerator devices\footnote{\href{https://coral.ai/products/}{https://coral.ai/products/}} for performing alternative experimental tests on the proposed implementations. These devices incorporate an Edge TPU for performing Machine Learning inferencing in existing systems.

Coral USB Accelerator works with Debian Linux, macOS and Windows 10. It supports TensorFlow Lite framework and is compatible with Raspberry Pi boards.

Mini PCIe, M.2 A+E key and M.2 B+M key Accelerators are PCIe devices that also enable the integration of the Edge TPU coprocessor into existing systems. These three devices support Debian Linux operating system and TensorFlow Lite framework.

The results obtained from experimental tests on systems equipped with these TPU-based devices will allow us to extend this study and quantify the performance improvement when using high speed serial interfaces between Edge TPUs and CPUs. This case will most likely provide an additional delay when loading data into the Edge TPU and result in similar effects to those described in equation \ref{eqn:su1}.

\section*{Acknowledgements}
Illustration A is provided by courtesy of Raspberry Pi Foundation. Illustration B is provided by courtesy of Coral.

\bibliographystyle{elsarticle-num} 
\bibliography{bibliography}






\end{document}